\documentstyle{elsart}
\begin{document}
\begin{frontmatter}
\title{Effect of shape anisotropy on transport in a 2-d computational
model: Numerical simulations showing experimental features observed
in biomembranes}

\author[Gauri]{Gauri R. Pradhan\thanksref{CSIR}}
\author[Gauri]{Sagar A. Pandit\thanksref{sagar-mail}\thanksref{sagar-add}}
\author[Gauri]{Anil D. Gangal\thanksref{DBT}}
\author[VS]{V. Sitaramam\thanksref{DBT}}

\address[Gauri]{Department of Physics, University of Pune, Pune 411 007, India}
\address[VS]{Department of Biotechnology, University of Pune, Pune 411
007, India}

\thanks[sagar-mail]{e-mail: sagar@prl.ernet.in}
\thanks[sagar-add]{Author for correspondence, Present address:
Physical Research Laboratory, Navrangpura,  Ahmedabad 380~009, India,
Ph. No: 91-79-6462129, Fax: 91-79-656502.}
\thanks[CSIR]{The work was partially funded by DBT(India) and CSIR(India)}
\thanks[DBT]{The work was funded by DBT(India)}

\begin{abstract}
 We propose a 2-d computational model-system
comprising a mixture of spheres and the objects of some other shapes,
interacting via the {\it Lennard-Jones} potential.  We propose a reliable
and efficient numerical algorithm to obtain void statistics. The
void distribution, in turn, determines the selective permeability
across the system and bears a remarkable similarity with features
reported in certain biological experiments.
\end{abstract}
\end{frontmatter}

\section{Introduction}
\label{intro}
The problem of packing of spheres plays a major role in the modeling of
many physical systems and has been studied for more than four 
decades. Some of the early examples~\cite{alder1,alder2,hoover1} of
the computer simulations of hard sphere liquids suggest the existence
of a first order freezing transition. The problem of
packing of spheres in two and three dimensions is of great
interest.  Recent investigations of such systems have focused on the
study of the statistical geometry of the dense sphere
packing. Such studies are important in the understanding of physical
properties of many systems, composed of a large number of
particles~\cite{speedy1,reiss1,torqu1,torqu2,rintoul1,reiss2,speedy2,sastri1,sastri2,sastri3}.

In this context we pose a question, with the motivation of studying the
transport across a two dimensional structure of packed circular disks
({\it membrane}), how does the packing change when the membrane is
doped with objects of various shapes and sizes (e.g. spheres
arranged rigidly in the form of rods of different lengths, L, T, X shapes
etc. See Fig. 1) ? In particular we investigate the effect of these
shapes on the distribution of {\it ``voids''}. The {\it``anisotropy''} in
 the interaction potential appears to play a key role in the
induction of large voids. 

As pointed out by Sastri et. al.~\cite{sastri1}, no 
algorithm is available to compute void statistics for the packing of
shapes other than spheres. In this paper we propose a simple numerical 
algorithm to compute void statistics. Unlike a probabilistic algorithm
(Monte Carlo), our algorithm is based on digitization and cell counting.

The paper is organized as follows. In Sec.~\ref{The-model-system}, we
describe the model system. A definition of ``void'' and an algorithm
to compute void statistics is given in Sec.~\ref{Voids}. The
results of numerical simulations and their relevance in lipid
biomembranes is discussed in Sec.~\ref{Results-and-Discussion} We
summarize the paper in Sec.~\ref{Summary}.  
\section{The model system}
\label{The-model-system}
 The configuration space of the model system (membrane)
is considered as a two dimensional space with periodic (toroidal) boundary
conditions. The constituents of the membrane are disks and dopants. 
\subsection{ The basic model }
We consider a membrane made up of only circular disks interacting
pairwise via the {\it Lennard-Jones} potential: 
\begin{eqnarray}
V_{LJ}(r_{ij}) = 4 \epsilon \sum_{i=1}^N \sum_{j = i+1}^N \Big( ({\sigma
\over r_{ij}} )^{12} - ( { \sigma \over r_{ij}})^6 \Big)\nonumber
\end{eqnarray}
where, $r_{ij}$ is the distance between the centers of the $i^{\rm
th}$ and $j^{\rm th}$ disks, $\sigma$ determines the range of hard
core part in the potential and $\epsilon$ signifies the depth of the
attractive part. We choose the number of disks such that
the area occupied by these disks is around $70\%$, which is less than
that of the close-packed structure but still large enough to produce
some closed voids.
\subsection { The model with impurities }
Further, we consider different {\it shape anisotropic} combinations
(dopants) consisting of $\kappa$ number of circular disks. We treat
each of these
combinations as a single rigid cluster. Several such dopants
(impurities) are considered. Fig. 1 shows some of these
impurities. The interaction between impurities and disks or other
impurities is obtained by superposing the {\it Lennard-Jones}
potential corresponding to each of the constituent disk in impurity.
We consider a membrane with circular disks and impurities amounting to
$10\%$ of the total number of circular disks, such that the area occupied
is still
$70\%$. 

These membranes are brought to an equilibrium configuration by the Monte
Carlo method~\cite{MC} at a fixed temperature. Fig. 2 and Fig. 3 show
typical
equilibrium configurations of membrane without and with impurities
respectively (The impurity in Fig. 3 is a rod shaped structure made up
of five disks (Rod$_5$), in general Rod$_\kappa$ for rod made up of
$\kappa$ number of disks). In the simulation the temperature is so chosen 
that
$k_B  T < 4 \epsilon$, where $k_B$ is the {\it Boltzmann}
constant. The equilibrium is confirmed by simulated annealing. 

\section{Voids and an algorithm for void statistics}
\label{Voids}
Now, we introduce the notion of an ``$r$-void'' in a membrane which is 
suitable for the description of transport across membrane and further,
propose an algorithm to compute statistical quantities such as
the number of voids in the membrane, the void size distribution
etc. 

We define an $r$-void as a closed area in a membrane devoid of disks or
impurities, and big enough to accommodate a circular disk of
radius $r$. Of course an $r$-void is also an $r^\prime$-void if
$r^\prime < r$.
\subsection{The algorithm to compute void statistics}
To compute the void statistics for $r$-voids, we increase the
radii of the disks forming the membrane (including the disks in the
impurities, without altering the positions of the centers) by an
amount $r$ (See Fig. 4). Then we digitize the entire membrane on
a suitably chosen grid. The choice of grid size depends on the
required accuracy and the typical sizes of the voids. The digitization
of circular disks is carried out by the Bressenham circle drawing
algorithm~\cite{Schaum}, modified to incorporate periodic boundary
conditions.  The number of voids in the membrane are
computed by flood filling~\cite{Schaum} every closed void with a different
color and then counting the number of colors. The sizes of various
voids can be obtained by counting the number of grid-cells filled by the
corresponding color. The termination of flood fill algorithm is
ensured since the voids are closed. In our case this condition is
automatically fulfilled in view of periodic boundary conditions.

 The geometric algorithms involving Vorenoi
polygons~\cite{sastri1,sastri2,sastri3} are mathematically satisfying
and are expected to be accurate but would take much more computation
time. On the other hand, as pointed in~\cite{sastri1}, the
probabilistic algorithm is time efficient but  requires a very large 
sample
size while dealing with small voids. 

Our algorithm is quite efficient as well as suitable even when
there are small voids in the membrane. We further note that the
algorithm can be easily generalized to higher dimensions. We expect
that the efficiency of this algorithm can be further enhanced by the use
of a multi-resolution adaptive grid.

\section{Results and Discussions}
\label{Results-and-Discussion}
The simulations were carried out for membranes of 
different compositions. Fig. 5 shows the graphs of the number of
$r$-voids as a function of $r$ measured in units of the radius of the constituent 
disks. Curve (a) shows void distribution in absence of impurities.
Curve (b) represents the void distribution in a membrane 
with rod shaped impurities made up of two disks  (Rod$_2$). 
 Curves (c) and (d) show the void distribution with L shaped
impurities made up of four disks (L$_4$) and rod like impurities made
up of four disks (Rod$_4$) respectively. It is clear from the graph 
that the number of large voids increases  with an increase in the
anisotropy of the impurity. Even though L$_4$ and Rod$_4$ occupy the
same area, Rod$_4$ being more anisotropic induces a larger number
of big voids than L$_4$. This fact can be clearly seen in Fig. 5, curves (c) and
(d). Moreover, the Fig. 2 and Fig. 3 demonstrate the fact that the
voids are mostly found in the neighborhood of the centers of anisotropy.
Further, to strengthen our claim that the shape anisotropy induces
voids, we compared two membranes. In one case  we added rod impurities
made up of two disks (Rod$_2$) in the assembly of circular disks, and
in the other case we added circular impurities of larger size, which
occupied the same area as that of Rod$_2$. We found that the former,
being more anisotropic, induced larger and more
numerous voids as compared to the later, though they occupied the same area. 

Thus, reduced to the bare essentials, the anisotropy in the
interaction potential of the constituents, is seen to be responsible for the
induction of large voids. If studied from the perspective of energy
minimization, as the potential becomes direction dependent, some
positions of the constituents are preferred over the other
positions. This induces large voids.

These features show a remarkable similarity with the observations reported 
in certain biological experiments~\cite{john}. These experiments deal
with the size-dependent permeation of non-electrolytes across
biological membranes. The effect of doping on the permeation of large
molecules was studied in these experiments. The liposome-membrane used
in these experiments was made up of mixture of two types of lipids
(cardiolipins and phosphatidylcholine) in a proportion 
1:10. The  understanding of the enhancement of transport in doped
membranes
needed an algorithmic statement. The ingredients at the algorithmic
level involved:
\begin{enumerate}
\item consideration of the structure as a strictly 2--dimensional assembly
\item the cross sections of molecules being considered as constituents
\item interactions of the constituents via the Lennard Jones potential 
\item permeating particles being considered as hard disks.
\end{enumerate}

The features reported in~\cite{john} bear a similarity with the
simulation carried out with Rod$_2$ as dopants. We have already seen
in numerical simulations (See Fig. 5, curves (a) and (b)) that the Rod$_2$
type of impurities induced large voids in the membrane. The appearance
of larger voids  naturally enhances the transport of large
particles. Thus an enhancement in the  transport of large
non-electrolytes like glucose, which was observed in the lipid mixture
~\cite{john} can possibly be understood using our simple approach.  

Further, apart from the biological implications, the model discussed is 
general enough to incorporate the studies of transport in various weakly 
bound granular media.
\section{Summary}
\label{Summary}

We have presented a numerical algorithm to compute the entire void
statistics in a two dimensional membrane consisting of circular disks
and dopants. We found that our simple two dimensional model has shown 
results consistent with features observed in a complex biological
system. The biological justification of the model and implications are 
discussed elsewhere~\cite{gauri}. Nevertheless, our model and
the proposed numerical algorithm which finds out the void statistics
in the model system are quite general and use no specific features of
any particular system. Therefore it is possible to use this method
effectively in various systems from diverse disciplines.  The result
that the shape anisotropy induces large voids in mixtures
may be used as a tool for achieving controlled selective permeability
across such a system by merely changing the shape of the constituents of
the mixture. 

{\bf Acknowledgments}
We thank N.V. Joshi, Deepak Dhar, H. E. Stanley  and S.S. Manna for fruitful
discussions.

\newpage
\noindent
{\bf Figure captions:}
\begin{description}
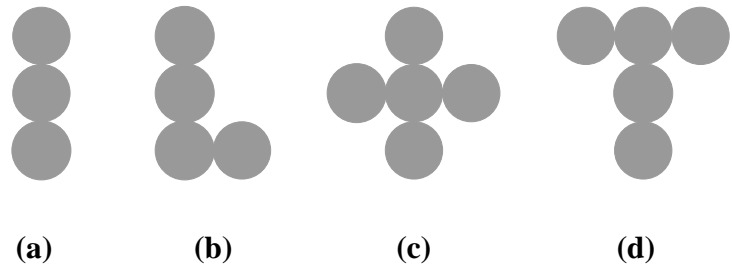
\item [Fig. 1] { Some examples of the impurities.
	\begin{description}
	\item[(a)] Rod type impurity made up of three circles
(Rod$_3$).
	\item[(b)] L type impurity made up of four circles (L$_4$).
	\item[(c)] X type impurity made up of five circles (X$_5$).
	\item[(d)] T type impurity made up of five circles (T$_5$).
	\end{description}}
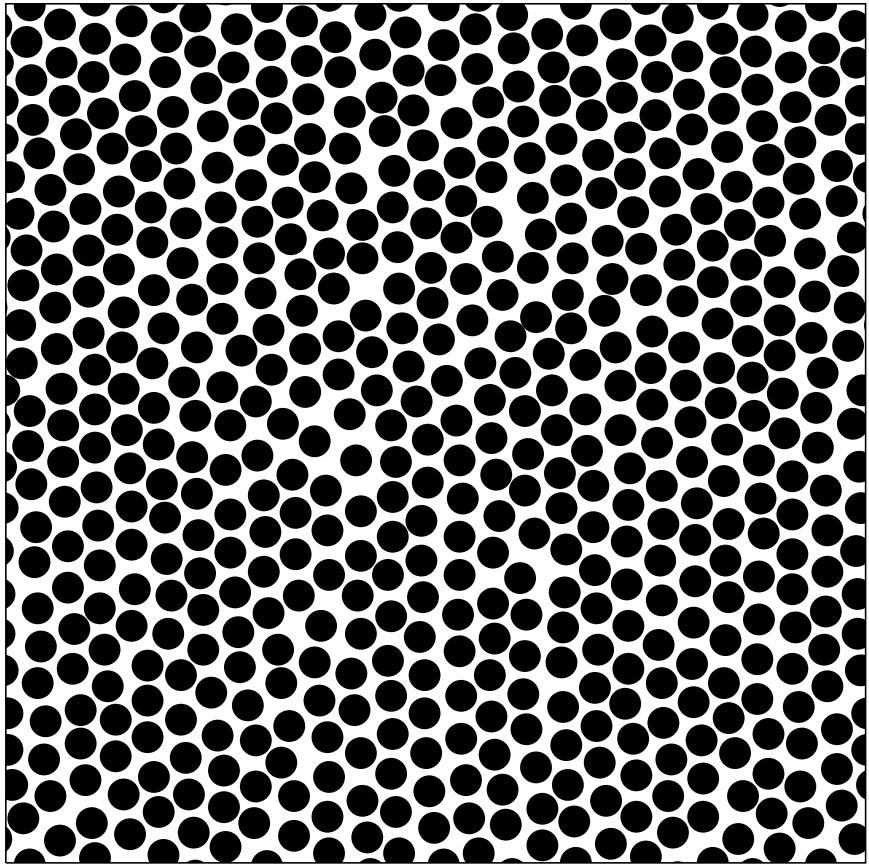
\item [Fig. 2] { Typical equilibrium configuration of a membrane
without impurity. There are 556 circular disks used to form this
membrane. The number is so chosen that the area occupied is $\approx
70\%$. The $\sigma$ in {\em Lennard-Jones} potential is chosen as two
times the radius of a circular disk.}
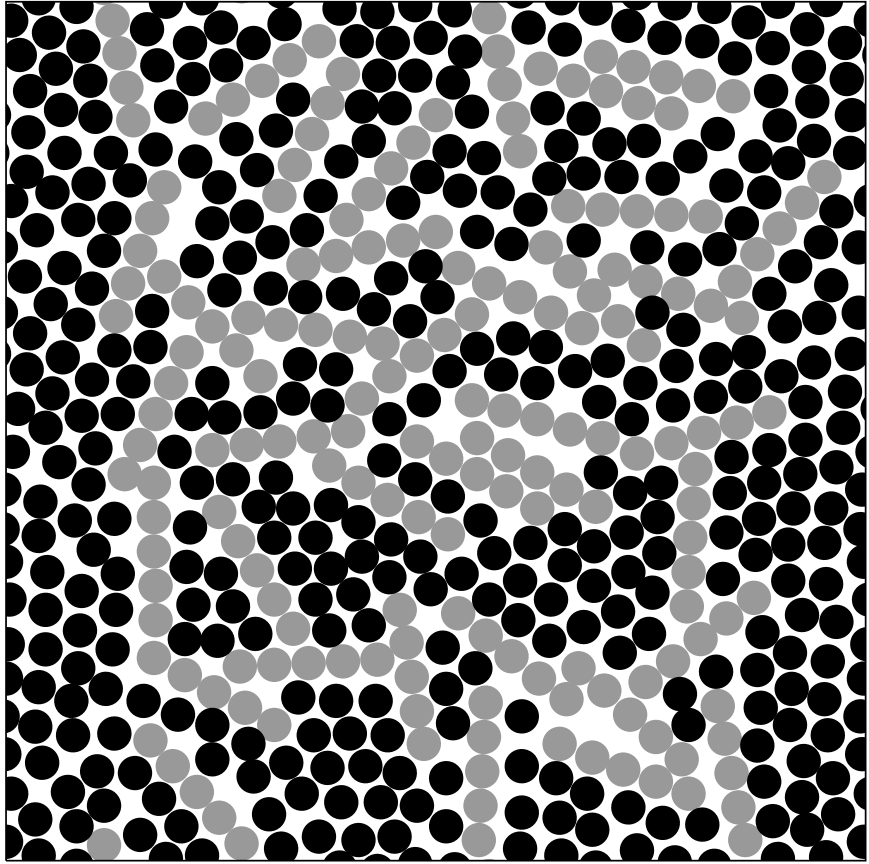
\item [Fig. 3] { Typical equilibrium configuration of a membrane
with impurity of type Rod$_5$. The amount of impurity is 1:10
proportion. All the other parameters are same as Fig. 2.}
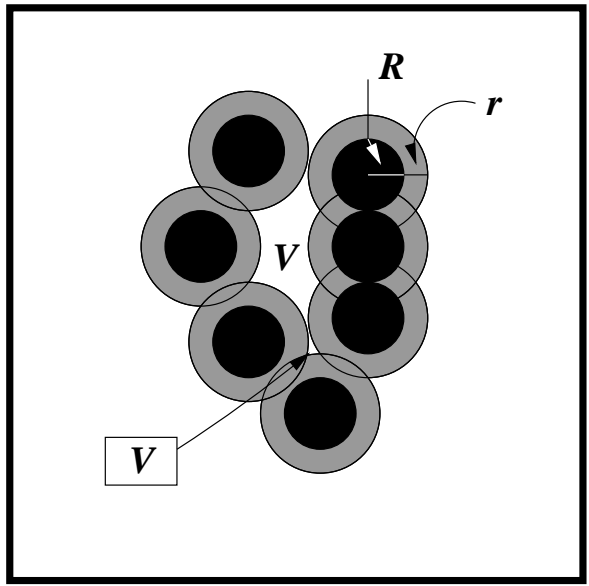
\item [Fig. 4] { Figure describes the algorithm to compute void
statistics. The radius of a circular disks (black disks) is $R$. These
disks are expanded by amount $r$, so that the region $V$ is the void
for particle of size $r$. }
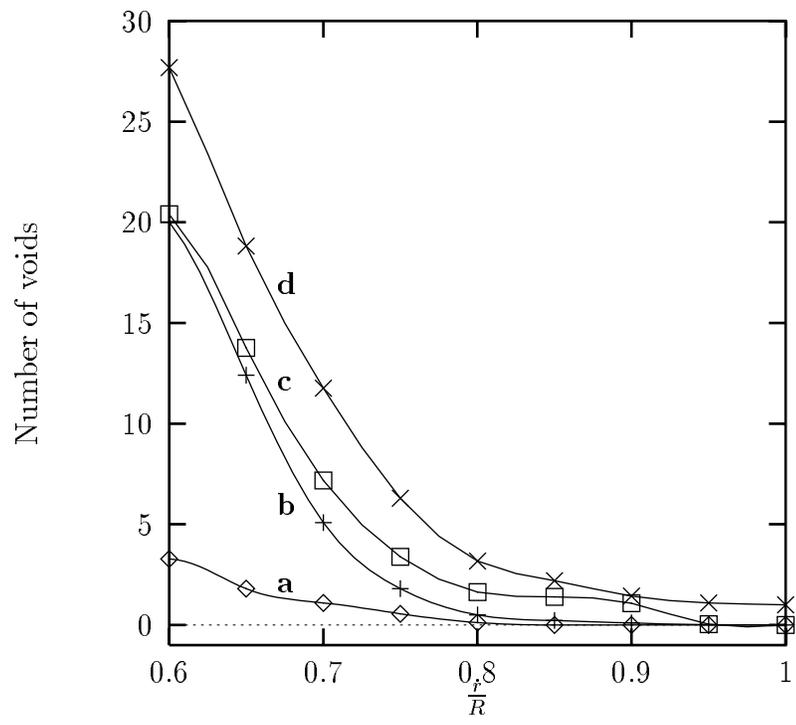
\item [Fig. 5] { The graphs of number of $r$-voids as a function of
$r$ measured in units of the radius of the constituents. 
\begin{description}
\item [Curve a] { The void distribution without impurities.}
\item [Curve b] { The void distribution with impurity of type Rod$_2$.}
\item [Curve c] { The void distribution with impurity of type L$_4$.}
\item [Curve d] { The void distribution with impurity of type
Rod$_4$.}
\end{description}
 Typically 10000 Monte Carlo steps are thrown away as
thermalisation, and it is ensured that total energy is minimized. The
curves are averaged over 100 Monte Carlo steps. } 
\end{description}
\end{document}